
\input phyzzx
\normalbaselineskip 14pt

\def\d{\partial}
\def\der#1#2{{\partial #1\over\partial #2}}
\def\eps{\epsilon}
\def\w{\sqrt{(E-E_1)\ldots (E-E_{2g+1})}}
\def\wtorus{\sqrt{(E-E_1)(E-E_2)(E-E_3)}}

\def\exp#1{{e^{\displaystyle #1}}}
\def\dsty{\displaystyle}
\def\ssty{\scriptstyle}
\def\abel{{\bf A}}
\def\bfn{{\bf n}}

\def\bfx{{\bf x}}
\def\bfy{{\bf y}}
\def\bfz{{\bf z}}
\def\bfs{{\bf s}}
\def\bft{{\bf t}}
\def\bfU{{\bf U}}
\def\bfV{{\bf V}}
\def\bfW{{\bf W}}
\def\bfS{{\bf S}}
\def\bfI{{\bf I}}
\def\bfE{{\bf E}}
\def\calL{{\cal L}}
\def\calO{{\cal O}}
\def\const{{\rm const}}
\def\dertwo{{\partial ^2\over\partial x^2}}

\def\psiad{{\psi^\dagger}}

\def\phiad{{\phi^\dagger}}
\def\one{{(1)}}
\def\two{{(2)}}

\def\bfznot{{{\bf z}_0}}
\def\<{\langle}
\def\>{\rangle}
\def\derivative{{\left.\der{}{\tau}\right|_0\psiad\psi(\tau)}}

\pubnum{91/26}
\titlepage
\title{
Non-linear WKB Analysis of the String Equation
}
\author{
F. Fucito
\foot{
email addresses: vaxtov::fucito and fucito@roma2.infn.it
}}
\address{
Dipartimento di Fisica, Universit\`a di Roma II {\it Tor Vergata}
and INFN, sezione di Roma {\it Tor Vergata},
Via Carnevale, 00173 Roma, Italy
}
\author{
A. Gamba
}
\address{
Dottorato in Matematica, Universit\`a di Milano,
via Saldini 50, 20133 Milano, Italy
}
\author{
M. Martellini
\foot{
permanent address:
Dipartimento di Fisica, Universit\`a di Milano, 20133
Milano, Italy and INFN, sezione di Pavia, 27100 Pavia, Italy;
email addresses: vaxmi::martellini and martellini@milano.infn.it
}
}
\address{
INFN, sezione di Roma I, piazzale Aldo Moro 5, Roma, Italy
}
\author{
O. Ragnisco
\foot{
email addresses: vaxrom::ragnisco and ragnisco@roma1.infn.it
}
}
\address{
Dipartimento di Fisica, Universit\`a di Roma I {\it La Sapienza}
and INFN, sezione di Roma I, piazzale Aldo Moro 5, Roma, Italy
}
\abstract
We apply non-linear WKB analysis to the study of the string equation.
Even though the solutions obtained with this method are not exact, they
approximate extremely well the true solutions, as we explicitly show
using numerical simulations.
``Physical'' solutions are seen to be separatrices corresponding
to degenerate Riemann surfaces. We obtain an analytic approximation
in excellent agreement with the numerical solution found by
Parisi {\it et al.} for the $k=3$ case.
\vfill
To appear in the proceedings of the {\it Research Conference
on Advanced Quantum Field Theory and Critical Phenomena}, held
in Como (Italy), June 17-21, 1991 -- World Scientific
\endpage
\pagenumber=1

\chapter{Introduction}
During the last two years, the partition function of two-dimensional
gravity has been set in correspondence with the $\tau$-function
of the KdV hierarchy, subjected to the constraint of the so-called
string equation
        \Ref\gromig{
        D.~Gross and A.~A.~Migdal
        \journal Phys. Rev. Lett. &64 (90) 127;
        \nextline
        E.~Br\'ezin and V.~A.~Kazakov
        \journal Phys. Lett. &236B (90) 144;
        \nextline
        M.~R.~Douglas and S.~H.~Shenker
        \journal Nucl. Phys. &B335 (90) 635.
        }.
In the case of pure gravity, described by
a matrix model with criticality index $k=2$, the string equation
reduces to the well known Painlev\'e type I equation for the
specific heat of the theory.
The string equation can be seen as a perturbation of a stationary
KdV equation, and thus solved in a semiclassical approximation
        \Ref\averagingmethod{
        I.~M.~Krichever
        \journal Funct. Analysis and Appl. &22 (88) 200.
        }
around the stationary KdV solution. Novikov and Krichever
        \Ref\novkrich{
        I.~M.~Krichever,
        {\it On Heisenberg Relations for the Ordinary
        Linear Differential Operators},
        to appear in Funct. Analysis and Appl.;
        \nextline
        S.~P.~Novikov
        \journal Funct. Analysis and Appl. &24 (90) 296.
        }
have conjectured
that such an approach lead to exact solutions already at
$0$-th order in perturbation theory. We have resorted to accurate
numerical simulations in order to check this conjecture.
On the other side, we show that ``physical'' solutions to the
$k$-th multicritical model are obtained as separatrices corresponding
to degenerate Riemann surfaces. We explicitly compare
the $k=3$ solution with the one found numerically by Parisi {\it et al.}
        \Ref\bmp{
        E.~Brezin, E.~Marinari and G.~Parisi
        \journal Phys.Lett. &242B (90) 35.
        }.
Purely asymptotic analysis seems to give in this case the same
degree of precision of the semiclassical approximation in the
non-degenerate case.

\section{The String Equation}

The string equation
$$
-x+\sum_{j=0}^{N}(j+1)\, t_{j+1} R_j[u(x)]=0
\eqn\stringeq
$$
(where $R_j$ are the Gel'fand-Dikii differential
polynomials)
for finite $N$ gives a constraint which is compatible with the first
$N$ flows of KdV. For $N\rightarrow\infty$ this gives a constraint
compatible with all of the KdV flows.
As a matter of fact, write the $j$-th KdV flow as
        $\der{u}{t_j}=K_j[u(x)]\equiv\der{}{x}R_j[u(x)]$,
differentiate \stringeq\ with respect to $x$ and compute
$$
\eqalign{
& \partder{}{t_s}\bigl(1+\sum_{j=0}^N (j+1)\, t_{j+1} K_j\bigr)=        \cr
& s\, K_{s-1}+\sum_{j=0}^N (j+1)\, t_{j+1} K'_j[K_s]=                   \cr
& s\, K_{s-1}+K'_s[\sum_{j=0}^N (j+1)\, t_{j+1} K_j]=                   \cr
& s\, K_{s-1}+K'_s[-1]=0                                                \cr
}
\eqn\stringcomp
$$
where
        $K'_j[\phi](u)\equiv\der{}{\eps}K_j(u+\eps\phi)|_{\eps=0}$
and we used known properties of the KdV flows, in particular
$(i)$
commutativity, {\it i.e.}
        $K'_j[K_s]-K'_s[K_j]\equiv [K_j,K_s]=0$;
$(ii)$
        $K'_s[1]=[K_s,1]=s\, K_{s-1}$,
expressing the fact that $1=\tau_{-1}$ is the first master symmetry
\Ref\mastersym{
        B.~Fuchsteiner
        \journal Progr. Theor. Phys. &70(83)1508;
        \nextline
        W. Oevel,
        {\it Mastersymmetries: weak action-angle structure for hamiltonian
        and non-hamiltonian dynamical systems},
        Paderborn preprint 1986
}
of the KdV hierarchy. We note by the way that \stringcomp\ can
be written in a more satisfactory invariant form
\Ref\magri{
        F.Magri, private communication
}
and easily generalized to arbitrary master-symmetries of the KdV equation.

\section{The Whitham method}

The Whitham method is the application to non-linear equations of the
semiclassical approximation known in the realm of linear equations as
WKB method.
Let us consider {\it e.g.} the wave equation
$$
\varphi_{xx}-{1\over c^2} \varphi_{tt}=0
\eqn\wave
$$
and the plane wave solutions $\varphi=A\exp{i(kx+\omega t)}$. If we are
studying propagation of light we have a ``small'' scale, {\it i.e.}
the characteristic length of oscillation, which is about $\epsilon=
10^{-6}$ times smaller than the natural unit length used by the
observer: so we can consider that solutions be locally given by
plane waves, while on the ``observational'' scale the parameters
$A,k$ are ``slowly'' varying:
$$
\dsty \varphi(x,t)=A(\eps x,\eps t)\,\exp{i {S(\eps x,\eps t) \over \eps}}
\eqn\wkb
$$
where
$\der{S}{X}=k(X,T)$,
$\der{S}{T}=\omega(X,T)$,
$X=\eps x$,
$T=\eps t$,
$\der{}{x}=\eps\der{}{X}$,
$\der{}{t}=\eps\der{}{T}$.
At 0-th order, we find the eikonal equation of geometrical optics:
$(\der{S}{X})^2={1\over c^2}(\der{S}{T})^2$.
The evolution of $k=\der{S}{X}$ gives the paths
of the ``rays''; the eikonal equation is equivalent to
$$
\cases{ k^2 ={\dsty\omega^2\over\dsty c^2} &\cr
                {\dsty \der{k}{T}  =\der{\omega}{X} }   &\cr}
\eqn\wkbdue
$$
In other words, we started with a class of exact solutions
$\varphi(x,t;A,k,\omega)=A \exp{i(kx+\omega t)}$
of \wave\ and passed from the precise description of the oscillation
process to the approximate description of the ``slow'' variation of
parameters $k,\omega$ given by \wkbdue; as a matter of fact we
averaged over the rapid variation of the function $\varphi$ and chose to
observe
only ``secular'' variations.
The same method could have been readily applied to non-linear equations
if we had
\item{\it i)}
a family of exact solutions depending on an adequate number
of parameters $E_1,\ldots, E_{2g+1}$;
\item{\it ii)}
a way of ``averaging out'' fast oscillations, in order to
obtain an analog of the second equation in \wkbdue,
which we will call ``Whitham equation''.

This is the case for the KdV equation
        \Ref\nlwkb{
        G.~B.~Whitham
        \journal Proc. Roy. Soc. &A283 (65) 238;
        \nextline
        H.~Flaschka, M.~G.~Forest and D.~W.~McLaughlin
        \journal Comm. Pure Appl. Math. &33 (80) 739;
        \nextline
        S.~Yu.~Dobrokhotov and V.~P.~Maslov
        \journal Russ. Math. Surv. &36 (81) 221.
        }
and more generally for equations that can be written in the form
        \foot{
                $[L,A]=LA-AL$
        is the usual commutator of differential operators, while
                $\der{}{y}$ and $\der{}{t}$
        are supposed to act on the coefficients
                $u_j, v_k$.
        }
$$
\der{L}{t}-\der{A}{y}+[L,A]=0,
\eqn\lax
$$
that is, as the compatibility condition
for the existence of a solution $\psi$ of the linear system
$$
\cases{
        L\psi ={\dsty \der{\psi}{y}}            &\cr
        A\psi ={\dsty \der{\psi}{t}}            &\cr
}
\eqn\kdvlinsys
$$
where
        $L=\sum_{j=0}^n u_j(x,y,t) {\d^j\over \d x^j}$,
        $A=\sum_{k=0}^m v_k(x,y,t) {\d^k\over \d x^k}$
are differential operators with scalar or matrix coefficients: as is
known, these equations admit large sets of exact solutions (the so-called
$g$-zone solutions) expressed in term of the Riemann $\theta$ function.

In particular, the KdV equation
$$
4u_t=u_{xxx}-6 u u_x
\eqn\kdv
$$
can be written as
        $L_t=[L,A]$
with
$$
\eqalign{
L       &=-\d^2+u(x,t)                                          \cr
A       &=\d^3-{3\over 2} u(x,t) \d - {3\over 4} u_x(x,t),      \qquad
                \d\equiv\ {\d\over\d x}                         \cr
}
\eqn\kdvlaxpair
$$
and admits the ``cnoidal wave'' solution
$$
u(x,t)=\wp(x-vt;g_2,g_3)
\eqn\cnoidal
$$
where $\wp$ is the Weierstrass elliptic function
and $g_2,g_3$ are arbitrary constants. Roughly speaking,
the Whitham method consists in slowly varying the constants $g_2,g_3$, that
is in finding the correct dependence $g_2=g_2(X,T)$, $g_3=
g_3(X,T)$, in order to approximate either ({\it i}) new solutions of
\kdv\ (corresponding {\it e.g.} to non-periodic initial data!) or
({\it ii}) solutions to the perturbed equation
$$
4u_t=u_{xxx}-6 u u_x +\eps K(x).
\eqn\pertkdv
$$
The cnoidal wave solution is a so-called $1$-zone ($g=1$) solution;
general $g$-zone solutions are given by
        $$
        u(x)=-2\der{^2}{x^2}\log\theta(\bfU x| B)+C
        \eqn\gzonesol
        $$
where the exact form of
        $\bfU=\bfU(E_j)$,
        $B=B(E_j)$,
        $C=C(E_j)$
is given in the Appendix.

\chapter{The Whitham method}

Let's start with perturbation
theory for a non-linear equation ({\it e.g.} the stationary KdV equation):
        $$
        u_{xxx}-12\, u u_x=\eps K
        \eqn\whithone
        $$
and look for solutions in the form
$$
u=u_0+\eps u_1+\eps^2 u_2+\cdots
\eqn\whithtwo
$$
with
$$
u_k=u_k(\bft|X)=u_k({\bfS(X)\over\eps}|E_j(X)),
\qquad X=\eps x,\qquad k=0,1,2,\ldots
\eqn\whiththree
$$
where ${1\over\eps}S_1(X),\ldots,{1\over\eps}S_{g}(X)$
are rapidly oscillating functions
that will be determined in the following, and $u_k$ depends on
some parameters $E_j=E_j(X)$
which on their turn are slowly varying with $x$,
and will be determined in the sequel.
Substitute \whithtwo\ and \whiththree\ in \whithone. Note that the form of
the functions $u_k$ implies
        $\der{}{x}=\der{\bfS}{X}\cdot\der{}{\bft}+\eps\der{}{X}$.
Now take the various order in $\eps$: {\it e.g.} at order $\calO(1)$ we get
$$
(\der{\bfS}{X}\cdot\der{}{\bft})^3 u_0(\bft|X) -
  12\, u_0(\bft|X)(\der{\bfS}{X}\cdot\der{}{\bft}) u_0(\bft|X) =0
\eqn\whithfour
$$
Let
${\cal L}=\der{\bfS}{X}\cdot\der{}{\bft}=
\sum_{j=1}^{g}\der{S_j}{X}\der{}{t_j}$,
and go on writing equations at all orders in the compact form:
$$
\eqalign{
\calO(1):\ \    \qquad\qquad    \calL^3 u_0-12\, u_0\, \calL u_0 &=0    \cr
\calO(\eps):    \qquad\qquad    \calL^3 u_1-12\, \calL(u_0u_1) &=F_1+K  \cr
\calO(\eps^k):  \qquad\qquad    \calL^3 u_k-12\, \calL(u_0u_k) &=F_k,
                \qquad          k=2,3,4,\ldots\cr                       \cr
}
\eqn\whithfive
$$
Note that only the first equation is non-linear, and that equations
for the order
$\calO(\eps^k)$,
$k=1,2,3,\ldots$,
differ only in the non-homogenous term
$F_k$. This term would not be present in na\"\i ve perturbation
theory, and comes from differentiation with respect to slow
variables: for instance
$u_{0x}=(\der{\bfS}{X}\cdot\der{}{\bft}+\eps\der{}{X})u_0$
produces the term
$\der{u_0}{X}$
of order $\eps$ giving contribution to $F_1$, together
with derivatives like $\der{}{x}S^j$ coming from $u_{0xx}$,
etc. (the explicit form of the term $F_1$ is given in the Appendix).
The idea is that at order $\calO(\eps)$ the term $F_1$ should
compensate for the perturbation term $K$ in order that the
``correction'' $u_1$ be bounded, and this gives equations for
the correct dependence of the parameters $E_j=E_j(X)$ on the
slow variables; otherwise, we can think of averaging
the $\calO(\eps)$ equation in
\whithfive\ over the fast variables $t_1,\ldots,t_{g}$,
thus remaining with the only variable $X$.
A third point of view is that the $\calO(\eps)$ equation has the form
${\bf L} u_1=F_1+K$, where ${\bf L}$ is a linear operator:
this means that periodic solutions exist iff $F_1+K$ is orthogonal to
${\rm Ker} ({\bf L^\dagger})$.
The three points of view are all equivalent and give
the same Whitham equations for $E_j=E_j(X)$.

Equation \whithfour\ has been written in explicit form in order to make
clear an important point: at any order the variable $X$
appears as a {\it parameter}, while equations are only in
the differential variables
        $t_1,\ldots,t_{g}$.
        \foot{
        This explains why the variables $x$ and $X$ are usually treated
        as independent variables in the two-scale method, and makes rigorous
        the usual argument of ``freezing'' the slow variable.
        }
Moreover, putting
$$
\der{\bfS}{X}=\bfU(E_j(X)),\qquad\bft=\bfU x
\eqn\whithsix
$$
we see that \whithfour\ becomes equivalent to the unperturbed
$u_{0xxx}-12u_0 u_{0x}=0$, so we immediately get the form of $u_0$
\foot{
        Actually, this is the complex form of $u_0$. A rigorous treatment needs
        the use of {\it real variables} $t_1,\ldots,t_{2g}$. In the $g=1$ case
        we would get
                $u_0(t_1,t_2)=\wp(2\omega\cdot t_1+2\omega'\cdot t_2;g_2,g_3)$.
}:
$$
u_0(\bft)=-2(\bfU\cdot\der{}{\bft})^2\log\theta(\bft|E_j)+C(E_j)
\eqn\whithseven
$$
that is, in the genus $g=1$ case,
$u_0(t_1)=\wp(2\omega\cdot t_1;\, g_2,g_3)$.

Functions of the form \whithseven\ are periodic in $t_1,\ldots,t_{g}$;
moreover, we have seen that the variables $\bft$ and X must be
regarded as independent: so we can average both sides of the
$\calO(\eps)$ equation
in \whithfive\ over $t_1,\ldots,t_{g}$, thus being left with the
only variable $X$. The equations for $E_1(X),\ldots,E_{2g+1}(X)$
thus obtained in the case of the stationary KdV equation have the form
$$
\der{}{X} {\ssty \sqrt{(E-E_1(X))(E-E_2(X))(E-E_3(X))}\, dE}=
-6 {
{\ssty E+r(E_1(X),E_2(X),E_3(X))}
\over
{\ssty \sqrt{(E-E_1(X))(E-E_2(X))(E-E_3(X))}}
}
\, {\ssty dE}
\eqn\whithamforstationarykdv
$$
where
$r(E_1,E_2,E_3)=\int_{E_3}^{E_2} {EdE\over\sqrt{(E-E_1)(E-E_2)(E-E_3)}} /
\int_{E_3}^{E_2} {dE\over\sqrt{(E-E_1)(E-E_2)(E-E_3)}}$
and $E$ is a dummy variable.

In general, the Whitham equations for KdV are differential conditions
on the functions
$p=p(E,E_j(X,T))$, $\Omega=\Omega(E,E_j(X,T))$ (quasi-momentum and
quasi-energy) appearing in the $E\rightarrow \infty$ leading term
$\psi\simeq \exp{px+\Omega t}$
of the solutions to the associated linear system \kdvlinsys.
For KdV they take the form
$$
\der{\Omega}{X}-\der{p}{T}=
{\langle\psiad K \psi\rangle\over\langle\psiad\psi\rangle}\der{p}{E}
\eqn\copy
$$
Equation \whithamforstationarykdv\ can equivalently be written as
$\der{\Omega}{X}=-6\der{p}{E}$.
A complete derivation and explanation of these formulas
is given in the Appendix.

\chapter{The Painlev\'e type I equation}

The Painlev\'e equation
$$
u''=6(u^2-x)
\eqn\painleveone
$$
(which is obtained from \stringeq\ when we leave only $t_3\neq0$)
can be differentiated once with respect to $x$ and seen as a perturbation
of the (integrable) stationary KdV equation:
$$
u'''=12 u u'+\eps K
\eqn\uffauno
$$
with $\eps=1,K=-6$.
An asymptotic relation between the two equations is obtained
through the change of variables
$$
\cases{
u(x)=\sqrt{\xi}\, v(\xi)        &\cr
\xi(x)={4\over 5} x^{5/4}.      &\cr
}
\eqn\uffadue
$$
It comes out
        \Ref\primo{
        F.~Fucito, A.~Gamba and M.~Martellini
        \journal Phys. Lett. &248B (90) 57.
        }
that solutions of \painleveone\ are asymptotic
to functions of the form
$u(x)=\wp({4\over 5}x^{5/4};12,g_3)$;
nothing is said about the parameter $g_3$. We will now resort to a
finer analysis of the problem.

We are here concerned
        \Ref\solitons{
        S.~P.~Novikov, S.~V.~Manakov, L.~P.~Pitaevskii and V.~E.~Zakharov,
        {\it Theory of Solitons},
        Plenum Publishing Company (New York, 1984).
        }
with the case of the torus of equation
        $(2w)^2=4E^3-g_2E-g_3=4(E-E_1)(E-E_2)(E-E_3)$;
we distinguish two cases; for
        $\Delta=g_2^3-27g_3^2>0$
we have real roots $E_1,E_2,E_3$ and two periods $2\omega$ and
$2\omega'$ which are respectively real and pure imaginary
        \Ref\abramowitz{
        H.~Abramowitz and I.~Stegun (ed.),
        {\it Handbook of Mathematical Functions},
        Dover (New York, 1972).
        }:
$$
2\omega=
2 \int_{E_3}^{E_2}{{\ssty dE}\over {\ssty \sqrt{4 E^3-g_2 E-g_3}}}=
        {{\ssty 2K(m)}\over{\ssty\sqrt{E_1-E_3}}},\qquad
2\omega'=
2 \int_{E_2}^{E_1}{{\ssty dE}\over {\ssty \sqrt{4 E^3-g_2 E-g_3}}}=
        {{\ssty 2iK(1-m)}\over {\ssty \sqrt{E_1-E_3}}},
\eqn\twentytwo
$$
where
$$
K(m)=\int_0^{\pi\over 2}{d\phi\over\sqrt{1-m\sin^2\phi}},\qquad
E(m)=\int_0^{\pi\over 2}\sqrt{1-m\sin^2\phi}\, d\phi,\qquad
m={E_2-E_3\over E_1-E_3}
\eqn\elliptic
$$
are standard elliptic integrals of the $1^{\rm st}$ and
$2^{\rm nd}$ kind, respectively, and $m$ is called the
Jacobi modulus.
We remind that the solutions of \uffauno\ with $\eps=0$
can be expressed in term of the Weierstrass
elliptic function (see Ref.~\abramowitz)
$$
\wp(x,g_2,g_3)=-\dertwo\log\theta_1({x\over 2\omega})-{\eta\over\omega}.
\eqn\twentythree
$$

The Whitham equations for \uffauno\ take the form
$$
\der{w}{X}dE=-6dp
\eqn\inthiscase
$$
or, explicitly,
$$
\der{}{X}\sqrt{4 E^3-g_2 E-g_3}\, dE\equiv
{\dsty{-\der{g_2}{X}}E-{\dsty\der{g_3}{X}} \over 2\sqrt{4 E^3-g_2E-g_3}} dE
=-6{E+r(X)\over\sqrt{4 E^3-g_2 E-g_3}}\, dE
\eqn\explicit
$$
giving
$$
\der{g_2}{X}=12,\qquad\der{g_3}{X}=12\, r(X).
\eqn\ten
$$
The first equation is readily integrated, giving $g_2=12\cdot X+\const$;
moreover, the second too is integrable by quadratures, because
$\der{w}{X} dE$ already has the same behaviour as $-6 dp$ for
$E\rightarrow\infty$,
so we need only to impose the normalization condition (compare with (5.4))
$$
{\rm Im}
\Bigl(\oint w\, dE\Bigr)=
{\rm Re}
\Bigl({1\over i}\int_{E_2}^{E_1}\sqrt{4 E^3-12X E-g_3(X)}\, dE\Bigr)=\const=h
\eqn\krich
$$
in order to get
${\rm Im}\oint\der{w}{X} dE=\der{}{X}{\rm Im}\oint wdE=0$
and $\der{}{X}wdE=-6dp$.
But \krich\ can be solved explicitly for $g_3$, yielding the correct
dependence $g_3=g_3(X)$.

In order to do this it is convenient to introduce new
parameters $(\lambda,m)$ in place of $(g_2,g_3)$:
$$
\eqalign{
&
\lambda={(E_1-E_3)^2\over 9},           \qquad
m={E_2-E_3\over E_1-E_3},                       \cr
&
g_2=12\lambda(1-m+m^2),                 \qquad
g_3=4\lambda^{3/2}(2-3m-3m^2+2m^3),             \cr
&
E_1=\sqrt{\lambda}(2-m),                \quad
E_2=\sqrt{\lambda}(-1+2m),              \quad
E_3=\sqrt{\lambda}(-1-m),                       \cr
&
\int_{E_2}^{E_1}\sqrt{4 E^3-g_2E-g_3}\, dE
        \equiv i{\lambda^{5/4}\over 5\sqrt{3}}\Phi(m);\cr
}
\eqn\eleven
$$
giving (for $\Delta>0$) the solution
$$
\lambda(m)=\left({5\sqrt{3}\over\Phi(m)}\right)^{4\over 5}.
\eqn\twelve
$$
In Fig. 1 we show the form of the resulting function $x=x(m)$,
both for the cases $\Delta>0$ and $\Delta<0$ (for graphic convenience
we plotted $x$ versus ${1\over m}$ instead of $m$). The pole
in the $\Delta<0$ region gives rise to two distinct curves
in the space of parameters
$(g_2,g_3)$, which we plotted in Fig.~2.
Knowing the properties
        \Ref\duval{
        P.~du~Val,
        {\it Elliptic Functions and Elliptic Curves},
        University Press (Cambridge, 1973)
        }
of the Weierstrass elliptic function $\wp$
we see that for $x={g_2\over 12}>0$ the Whitham
method gives us either small oscillations (curve in the $\Delta>0$
region) or a sequence of poles (lower part of the curve, lying in the
$\Delta<0$ region).
If we start with oscillatory behaviour at $+\infty$ and go in
the direction of decreasing $x$, we ultimately reach $\Delta=0$
where a transition to polar behaviour occurs.
Note that for $x<0$ we can have only polar behaviour. The two
parts of the curve correspond to solutions having or not
having poles at $+\infty$.
The ``physical'' solutions will be seen to be separatrices lying between
these two kinds of solutions.

The function $\Phi(m)$ is computed reducing the integral in \krich\
to standard elliptic integrals
        \Ref\bateman{
        H.~Bateman,
        {\it Higher Transcendental Functions},
        McGraw Hill (London, 1954)
        }.
We get
        $\int_{E_3}^{E_2}{dE\over 2w}={K(m)\over \sqrt{3} \lambda^{1/4}}$,
        $\int_{E_2}^{E_1}{dE\over 2w}={K(1-m) \sqrt{3} \lambda^{1/4}}$,
        $\int_{E_3}^{E_2}{EdE\over 2w}={\lambda^{1/4}\over \sqrt{3}}{\ssty
K(m)}
        -{\ssty \sqrt{3}\lambda^{1/4} E(m)}$,
        $\int_{E_2}^{E_1}{EdE\over 2w}={\lambda^{1/4}(-1-m)\over \sqrt{3}}
        i {\ssty K(1-m)}+{\ssty \sqrt{3}\lambda^{1/4}i E(1-m)}$,
and finally
$$
\eqalign{
\int_{E_3}^{E_2}\sqrt{4 E^3-g_2 E-g_3}
        &={2\over 5}{\lambda^{5/4}\over\sqrt{3}}
        (18(2-3m+m^2)K(m)-36(1-m+m^2)E(m)),     \cr
\int_{E_2}^{E_1}\sqrt{4 E^3-g_2 E-g_3}
        &={2\over 5}{i\lambda^{5/4}\over\sqrt{3}}
        (-18(m+m^2)K(1-m)+36(1-m+m^2)E(1-m)),   \cr
\Phi(m)
        &=-18(m+m^2)K(1-m)+36(1-m+m^2)E(1-m).   \cr}
\eqn\sixteen
$$
This procedure gives us also the ``actions'' $S_1,S_2$:
$$
S_1=-{1\over 3}\int_{E_2}^{E_1}\sqrt{4 E^3-g_2 E-g_3}\, dE,\qquad
S_2=-{1\over 3}\int_{E_3}^{E_2}\sqrt{4 E^3-g_2 E-g_3}\, dE,
\eqn\seventeen
$$
because from $\der{w}{X}dE=-6dp$ it follows
$$
\der{S_1}{X}=U_1=2\int_{E_2}^{E_1}dp,\qquad
\der{S_2}{X}=U_2=-2\int_{E_3}^{E_2}dp.
\eqn\eighteen
$$
Take $S=S_1+\tau S_2$ as prescribed by
\whithsix, (5.15) and (5.16),
where
$\tau=i{K(1-m)\over K(m)}$ is the 1-dimensional analog of the
period matrix $B$, and find
from \sixteen\ and
the Legendre relation
$EK'+E'K-KK'={\pi\over 2}$
(see Ref.~\abramowitz):
$$
S=i{\sqrt{3}\over 5}\lambda^{5/4}\cdot
        {12(1-m+m^2)\over K(m)}={4\over 5}i {x\over 2\omega}.
\eqn\nineteen
$$
The Whitham solution has the form
        $$
u_0(x)= -\wp({4\over 5} i x+\omega;12 x,4 I(x) x^{3/2})=
        -\sqrt{x}\wp({4\over 5} i x ^{5/4}+\omega;12,4 I(x));
\eqn\twenty
$$
(We have used here the homogeneity property
$\wp(tx;{g_2\over t^4},{g_3\over t^6})$
$={1\over t^2}\wp(x;g_2,g_3)$, $t=x^{-1/4}$;
the $\omega$-shift is needed in order to get non-singular solutions).
We thus recover the
``Boutroux'' asymptotic form \uffadue,
but with a more precise
$$
I(x)=I(m(x))={2-3m-3m^2+2m^3\over(1-m+m^2)^{3/2}}.
\eqn\twentyone
$$
Asymptotic analysis
        \Ref\ben{
        C.~Bender and S.~Orszag,
        {\it Advanced Mathematical Methods for Scientist and Engineers},
        McGraw-Hill (London, 1978).
        }
shows that the amplitude of the oscillatory solutions of \painleveone\
decreases at $+\infty$ as ${1\over x^{1/8}}$. This feature, not
recovered by the simple Boutroux-type solution, is obtained from
the fine tuning realized by the term $I(x)$. Just use the expansion
$$
K(m)={\pi\over 2}(1+{m\over 4}+{9\over 64} m^2+\cdots),\qquad
E(m)={\pi\over 2}(1-{m\over 4}-{3\over 64} m^2+\cdots),
\eqn\expansion
$$
for $m\rightarrow 0$ (see Ref. \abramowitz).
{}From \eleven, in the limit $m\rightarrow 1$,
we get
$x={g_2\over 12}\simeq\lambda(m)$,
$\Phi(m)\simeq{135\pi\over 8}(1-m)^2$,
$\lambda(m)\simeq({8\sqrt{3}h\over 27\pi})^{4/5}(1-m)^{-8/5}$,
$-E_1,-E_2\simeq\sqrt{x}$,
$E_1-E_2\simeq\sqrt{8h\over \sqrt{3}\pi} x^{-1/8}$
(Notice that the function $\wp(ix+\omega)$ oscillates
between the extremal
values $-E_1$ and $-E_2$, see Ref.~\duval).

If we want to examine the case $x<0$, corresponding to $\Delta<0$,
formulas \eleven\ are no longer convenient, and we resort to the
following real parameterization
$$
m_*={1\over m}={1\over 2}+i\sigma,\quad
\lambda_*=m^2\lambda=-\rho^2,\quad
i\lambda^{5/4}\Phi(m)=-i\lambda_*^{5/4}\Phi(m_*)
\eqn\newpar
$$
The solution corresponding to \twelve\ in the complex case becomes
$$
\lambda_*(m_*)=-\left({5\sqrt{3}h\over{\rm
Re}(\sqrt{i}\Phi(m_*))}\right)^{4/5}.
\eqn\complexsolution
$$
A numerical computation shows that the function
$\Phi(m_*)\equiv\Phi({1\over 2}+i\sigma)\equiv\Phi(\sigma)$
has a zero for
$\sigma=\sigma_0=-0.231026398427\ldots$.

In the case $\Delta<0$ real and pure imaginary combinations of periods
are given by
$\omega={K(\tilde m)\over\sqrt{H}}$,
$\omega'_2={iK(1-\tilde m)\over\sqrt{H}}$, where
$H^2=3E_1^2-{g_2\over 4}=|{3-4\sigma^2\over1+4\sigma^2}|$,
$\tilde m={1\over 2}+{\sigma\over\sqrt{1+4\sigma^2}}$
(see Ref.~\abramowitz).
With this parameterization we recover the numerical result
of Ref.~\ben, asserting that the distance of poles goes asymptotically
as $\tilde c\over x^{1/4}$, with $\tilde c=7.276726\ldots$; as a matter
of fact we find
$c={2\omega'_2\over i}={2\over\sqrt{3}}
K({1\over 2}-{\sigma_0\over\sqrt{1+4\sigma_0^2}})
({3-4\sigma_0^2\over 1+4\sigma_0^2})^{1/4}=2.970711275212\ldots$,
which exactly coincides with the result of Ref.~\ben\ after the
rescaling $\tilde c=\sqrt{6}\, c$
(due to our factor $6$ in \painleveone).
An analogous reasoning for $x>0$ gives the period of the
oscillatory solutions going as $c\over x^{1/4}$, with
$c={2\omega'\over i}={\pi\over\sqrt{3}}$.

In Fig. 3 we show the approximate solution
(with $h=1$), together with an exact numerical solution obtained
with the Runge-Kutta method
(the Painlev\'e equation is satisfied with an error of $10^{-14}$).
Fig. 4 is a magnification of the region
around the zero where the approximation seems to be less effective. We see that
the solution we are considering seems to be out of phase with respect to the
exact numerical one. This is no surprise
because we really have neglected a phase: the Whitham method made variating
the ``constants'' $g_2$ and $g_3$, but in $\wp(x+c;g_2,g_3)$ we have
also a third integration constant $c$, that we assumed to be zero.
We do believe that an equation for this phase can be deduced
        \Ref\fas{
        I.~M.~Krichever
        \journal Sov. Math. Dokl. &27 (83) 757.
        },
and its behaviour will be discussed in a forthcoming paper.
Here we have just made a fit of such a
phase on the side of the positive $x$, getting a correction
$\delta\phi(x)\simeq{.175\over(x-.0098)^{.538}}$.
We have also done the same on the negative $x$ side but we got
a really tiny correction that we chose to neglect.
The fit of the ``experimental" data is shown in
Fig. 5, while Fig. 6 shows the effect of putting in the correction
by hand (for $x\simeq 0$ we let $\delta\phi(x)$ die smoothly).

\chapter{Degenerate solutions}

The solutions $u(x)$ of Painlev\'e-like equations represent
specific heats of the random matrix models, which in the
planar limit must satisfy the boundary condition
        $u(x)\simeq x^{1/k}$
($x\rightarrow\infty$)
for scaling arguments of the partition function:
        $Z\sim x^{-\gamma+2}$,
where $\gamma$ is the string susceptibility. Thus
        $u(x)\equiv\der{^2F}{x^2}\sim\der{}{x} x^{-\gamma+1}\sim x^{-\gamma}$,
and the result comes from the fact that $\gamma={1\over k}$
in the proximity of the critical point.
The situation is common to other non-linear physical models in the
critical r\'egime. It is well-known that the problems of mathematical
physics must be complemented by boundary conditions,
and that the boundary conditions contain in some sense the physics of
the problem. In our particular case we come to the request that
the solutions to the string equation for the $k$-th multicritical
model (which is obtained from \stringeq\ putting all $t_j=0$,
except $t_{k+1}$) must satisfy the
physical constraint $u(x)\simeq x^{1/k}$ for $x\rightarrow +\infty$.
However, the Whitham method gave us either small oscillations modulating
over $-\sqrt{x}$ (see Fig. 3), or solutions with poles.
The only possibility to get non-periodic solutions is to considerate
degenerate Riemann surfaces, where the length of the bands is sent to $0$.
These solutions are degenerate cases of the periodic solutions and are
themselves unstable separatrices, lying between the two sets of solutions
with poles and without poles for $x\rightarrow +\infty$.

All we have to do is to compute the spectral curve corresponding
to the given $k$-th stationary KdV equation and imposing the
coincidence of the pairs of branch points; $g=k-1$ conditions
are found by requesting that $dw=dE^{g+{1\over 2}}+\calO(1)$;
one more condition comes from fixing the periods of the solutions
at $x=\pm\infty$; the last $g$ conditions come from the request
that the branch points coalesce in pairs.

For the $k=2$ case we require $dw\simeq dE^{3/2}+\calO(1)$, giving
$$
\cases{
E_1+E_2+E_3=0           &\cr
E_1E_2+E_2E_3+E_3E_1=c. &\cr
}
\eqn\kappadue
$$
The asymptotic condition fixes
$c=3$, and we ask for $E_2=E_3$. This gives $E_2=E_3=-{E_1\over 2}=1$.
The $\theta$ function degenerate to a combination of
hyperbolic functions
and we finally get
$$
u_0(x)=\cases{
\sqrt{x}\, (1-{3 \over (\cosh(\sqrt{3}\cdot{4 \over 5}x^{5 \over 4}))^2}),
& $x\geq 0$, \cr
\sqrt{-x}\, (\wp({4 \over 5}(-x)^{5 \over 4};-12,-I({1 \over 2}+i\sigma_0)),
& $x< 0$. \cr}
\eqn\tuttto
$$
(In the $x\leq 0$ case the surface does not degenerate, but
$I(x)\rightarrow I({1\over 2}+i\sigma_0)=\const$, where $\sigma_0$
is the constant introduced in the previous section).
The function $u(x)$ is plotted in Fig. 7.
In $x=0$ we get a cusp as we are trying to connect at finite $x$
two asymptotic solutions: a smoother curve would probably
require a phase-type correction as suggested in the previous section.

In the $k=3$ case the condition $dw=dE^{5/2}+\calO(1)$
gives
$$
\cases{
\sum_{i=1}^5 E_i =0             &\cr
\sum_{i<j} E_i E_j =0           &\cr
\sum_{i<j<k} E_i E_j E_k = c.   &\cr
}
$$
Asymptotic conditions fix $c=\pm{5\over 8}$ for $x\rightarrow\pm\infty$;
we put $E_2=E_3=s$, $E_3=E_4=t$, and find
$s=\pm({1+i\sqrt{5}\over 4})$, $t=\bar s$, $E_1=\pm 1$
($x\rightarrow\pm\infty$).
Again the $\theta$ function factorizes in products of trigonometric
and hyperbolic functions (for details see
        \Ref\nostro{
        F.~Fucito, A.~Gamba and M.~Martellini,
        {\it On the String Equation and the Whitham Method},
        Rome preprint ROM2F -- 91/14
        }
and
        \Ref\fay{
        J.~D.~Fay,
         {\it Theta functions on Riemann Surfaces},
        Lecture Notes in Math. {\bf 352},
        Springer (Berlin, 1973)
        })
giving
$$
v_0(x)=\cases{
1-2a^2
{(\coeff{b}{a}-\coeff{a}{b})\sinh (ax+2\ln \coeff{a}{b})\sin (bx)
  +2\cosh (ax+2\ln \coeff{a}{b})\cos(bx)-2 \over
 ({\dsty\sinh(ax+2\ln}\coeff{a}{b})-\coeff{a}{b}\sin (bx))^2}, & $x< 0$; \cr
-1-2b^2
{(\coeff{a}{b}-\coeff{b}{a})\sinh (bx+2\ln \coeff{b}{a})\sin (ax)
  +2\cosh (bx+2\ln \coeff{b}{a})\cos(ax)-2 \over
  ({\dsty\sinh(bx+2\ln }\coeff{b}{a})-\coeff{b}{a}\sin (ax))^2},&$x\geq 0$.\cr}
\eqn\pizzauno
$$
where $a=(30)^{1 \over 4}\cos{\vartheta \over 2},
b=(30)^{1 \over 4}\sin{\vartheta \over 2}, \vartheta=\arctan
{1 \over \sqrt{5}}$.
(We checked that this solution satisfies the higher order
stationary KdV equation).
Fig. 8 shows that
        $u_0(x)={}^3\sqrt{x}\cdot v_0({6\over 7}x^{7\over 6})$
 approximate very well the form of the
solution found numerically in Ref. \bmp,
except that in the proximity of $x=0$.
(In the graph we have shifted
$z_1\mapsto z_1+{1 \over 4}, z_2\mapsto z_2 -{1 \over 4}$
for $x<0$ to match the phase of
Ref \bmp. The resulting function still satisfies the KdV equation).

\ack

One of us (A.~G.) wants to thank I.~Krichever for having clarified to him
several parts of his work, and F.~Magri for useful discussions.

\chapter{Appendix}

In sections 1, 2, 3 we review some facts
about algebraic geometry and KdV equations, mainly
for notational convenience. In section 4 we report the proof
of Krichever's theorem, following Ref.~\averagingmethod.

\section{Complex curves}
The algebraic equation
$$
w^2=E^{2g+1}+a_1 E^{2g-1}+a_2 E^{2g-2}+\ldots +a_{2g-1} E+a_{2g}
        \equiv p_{2g+1}(E)
\eqn\curve
$$
defines a curve $\Gamma$ in the complex plane of the variables
$(E,w)$.
The curve is compactified at the $\infty$ and is known to be
topologically equivalent to a compact surface with $g$ holes.
If the polynomial $p_{2g+1}(E)$ has $2g+1$ real roots $E_1,\ldots,E_{2g+1}$,
we can draw them on the complex plane and use solid lines for the
segments $(E_{2k-1},E_{2k})$, where the square root
$
w=\pm\sqrt{p_{2g+1}(E)}=\pm\sqrt{(E-E_1)(E-E_2)\cdots (E-E_{2g+1})}
$
takes real values. In spectral theory these are the forbidden zones
of the spectrum.

Coordinates on $\Gamma$ are given by
$$
\cases  {u=E                    &       almost everywhere,              \cr
         u=\sqrt{E-E_j}         &       in the neighborhood of $E_j$,   \cr
         u={1\over\sqrt{E}}     &       in the neighborhood of $\infty$.\cr}
\eqn\coordinates
$$
Consider integrals of the form
        $\int \Omega_k=\int {E^k dE\over 2\w}$,
        $k=0,1,2,\ldots$
Using  \coordinates\ it is easy to see that
$\Omega_0,\ldots,\Omega_{g-1}$
are everywhere non-singular, while
$\Omega_g,\Omega_{g+1},\ldots$ have
{\it poles} at the infinity of order $2,4,\ldots$, etc.

To fix a basis of differentials we chose first a canonical
        \Ref\dubrovin{
        B.~A.~Dubrovin
        \journal Russ. Math. Surv. &36:2 (81) 83.
        }
basis of
paths $a_1,\ldots,a_g$ and
$b_1,\ldots,b_g$ on $\Gamma$ and take
$\omega_1,\ldots,\omega_g$
as linear combinations of
$\Omega_0,\ldots,\Omega_{g-1}$ satisfying the normalization condition
$$
\oint_{a_k} \omega_j\equiv
2\int_{E_{2k-1}}^{E_{2k}} \omega_k=\delta_{jk},
\qquad j,k=1,\ldots,g
\eqn\normalizationone
$$
We are left with the $b$-periods, forming a $g\times g$ matrix
$
B_{jk}=B_{kj}=\oint_{b_k}\omega_j\equiv
2\int_{E_{2k}}^{E_{2g+1}} \omega_j.
$
Differentials with poles of order $2j$ will be indicated by
$\omega^{(2j-1)}$ for future commodity, and can be fixed by requiring that
they go at the infinity as
$
\omega^{(2j-1)}\simeq dE^{j-{1\over 2}} + {\cal O}(1).
$
the arbitrariness on the holomorphic tail
can be eliminated by imposing
the $2g$ real conditions
$$
        {\rm Im}\oint_{a_k}\omega^{(2j-1)}=0,\quad
        {\rm Im}\oint_{b_k}\omega^{(2j-1)}=0, \quad k=1,\ldots,g.
\eqn\normalizationtwo
$$
(A\-no\-ther stan\-dard choice of the nor\-ma\-li\-za\-tion is to im\-po\-se
instead of \normalizationtwo\ the $g$ com\-plex con\-di\-tions
        $\oint_{a_k} \omega^{(2j-1)}=0$,
        $k=1,\ldots,g$).
We will also use the notation
$$
dp=\omega^{(1)}\simeq d\sqrt{E},\qquad
d\Omega=\omega^{(3)}\simeq dE^{3/2}, \qquad
E\rightarrow\infty;
\eqn\newnotation
$$
these are the differentials of the quasi-momentum $p(E)$ and
quasi-energy $\Omega(E)$, fundamental in the theory of the KdV
equation (see 5.11);
$p(E)=\int_\infty^Edp$ and
$\Omega(E)=\int_\infty^Ed\Omega$
are {\it multivalued functions} with periods
$U_j=\oint_{b_j}dp$, $W_j=\oint_{b_j}d\Omega$, $j=1,\ldots,g$.
Note that $p(E)$ and $\Omega(E)$ are
{\it uniquely determined} by the asymptotic behaviour
and the normalization conditions.

        \section{Functions on the surface $\Gamma$}

The Abel map
$P\mapsto \abel(P)\equiv
        {}^t(\int_\infty^P \omega_1,\ldots,\int_\infty^P \omega_g)$
maps any point $P$ on $\Gamma$ on the $g$-dimensional torus
$C^g/\{\rm period\ lattice\}$.
The Abel map can be inverted by means of the Fourier series
$$
\theta(\bfz|B)=
\sum_{\bfn\in\zeta^g} e^{\pi i \bfn\cdot B\cdot \bfn+2\pi i \bfn\cdot\bfz }
\eqn\riemanntheta
$$
defining (for positive definite ${\rm Im} B$) the Riemann $\theta$ function,
which has the periodicity properties
$\theta(\bfz+{\bf e}_j)=\theta(\bfz)$,
$\theta(\bfz+B{\bf e}_j)=e^{-\pi i B_{jj}-2\pi i z_j}\theta(\bfz)$
(see Ref. \dubrovin).
As a matter of fact, a theorem of Jacobi asserts that the function
$$
f(P;P_1,\ldots,P_g)=\theta(\int_\infty^P\omega_k-
        \sum_{j=1}^g\int_\infty^{P_j}\omega_k+K_k)
\eqn\effe
$$
(where ${\bf K}=(K_k)$ is a certain constant vector) has exactly $g$ zeroes
$P_1,\ldots,P_g$.
This theorem gives an analog of the development of a rational
function in simple fractions.

\section{The  KdV equation}

The KdV equation
        $4u_t=u_{xxx}-6 u u_x$
admits the Lax representation
        $L_t=[L,A]$,
with
        $L$, $A$,
given by \kdvlaxpair, and has exact solutions of the form
$$
u(x,t)=-2{\d^2\over\d x^2}\log\theta(\bfU x+\bfW t+{\bf z}_0|B)+C(B)
\eqn\kdvsolution
$$
where (see \newnotation)
$$
U_j=\oint_{b_j}dp,\qquad
W_j=\oint_{b_j}d\Omega,\qquad
j=1,\ldots,g.
\eqn\wavevectors
$$
This can be seen as follows
        \Ref\krichsettesette{
        I.~M.~Krichever
        \journal Funct. Analysis. and Appl. &11 (77) 12.
        }.
The equation $L_t=[L,A]$ is the compatibility
condition for the system of linear equations
$$
\cases{
        L\psi=E\psi                     &\cr
        A\psi={\dsty\der{\psi}{t}}.     &\cr
}
\eqn\linearsystem
$$
For $E\simeq\infty$ we get $L\simeq\-d^2$, $A\simeq\d^3$, so the
asymptotic form of the common eigenvectors $\psi$ will be
$$
\psi(x,t;E)=\exp{p(E) x + \Omega(E) t}\cdot\phi(x,t;E),\qquad
{\rm  with}\quad p(E)\simeq\sqrt{E},\quad
                  \Omega(E)\simeq E^{3/2}\quad
{\rm as}\quad E\mapsto\infty.
\eqn\eigenvector
$$
The exact form of $\phi(x,t)$ is (compare with \effe; see Ref. \solitons)
$$
\phi(x,t;E)=
{\theta(\int_\infty^E\omega_k-\sum_{j=1}^g\int_\infty^{E(P_j)}\omega_k+
                                                U_k x+V_k t+K_k)\over
\theta(\int_\infty^E\omega_k-\sum_{j=1}^g\int_\infty^{E(P_j)}\omega_k+K_k)}
\eqn\bakerakhiezer
$$
for given $\Gamma$ and $P_1,\ldots,P_g$ on $\Gamma$.
It is easy to verify that
$\psi=\exp{px+\Omega t}\cdot\phi$\
is a one-valued function of $E$.
The function~$\psi$ is
{\it uniquely determined} by the behaviour at infinity
($\psi\simeq e^{E^{1/2} x+E^{3/2} t}$)
and the position of the $g$ poles $P_1,\ldots,P_g$,
as can be easily seen with the help of the Riemann-Roch theorem.
For $E\equiv k^2\rightarrow\infty$ we get
$$
\psi(x,t;E)=c \exp{kx+k^3 t}\cdot (1+{\xi_1(x,t)\over k}+
                             {\xi_2(x,t)\over k^2}+\cdots)
\eqn\asympt
$$
If $\psi$ satisfies $L\psi=E\psi, A\psi=\der{\psi}{t}$, then we can
collect terms of the same order in ${1\over k}$ and
$\xi_1, \xi_2,\ldots$ should satisfy some equation at any order.
It is easily seen that the
first of these  equation gives $u(x,t)=2\der{\xi_1}{x}(x,t)$, so for
this choice of the potential $u$ we get \linearsystem\ verified
at order ${\cal O}({1\over k})$. But note that $(L-E)\psi$, $(A-\der{}{t})\psi$
again have the behaviour $\exp{kx+k^3 t}$ and poles at $P_1,\ldots,P_g$, so
they must again have the form \asympt\ with no $\calO(1)$ term:
this means that they are identically
zero; so \eigenvector\ and \bakerakhiezer\ give
an exact solution to \linearsystem.
Developing \bakerakhiezer\ at $1^{\rm st}$ order in ${1\over k}$ we get
precisely formula \kdvsolution\ for
$u=2\der{}{x}\xi_1(x,t)$.

\section{Krichever's theorem}

We will here report the form found by Krichever for
the Whitham equations of systems of KP type, following
Ref.~\averagingmethod.
It is convenient to consider directly the general case of equations
of the form \lax,
as for instance the KP equation
$
3u_{yy}+\der{}{x} (4 u_t-6u u_x+ u_{xxx})=0
$.
\lax\ is the compatibility condition for the existence
of a solution $\psi$ of the linear system
        $L\psi=\der{\psi}{y}$,
        $A\psi=\der{\psi}{t}$.
The common eigenvector $\psi$ will be given here using a particular
{\it real} normalization, necessary for the successive averaging
procedure:
$$
\psi(x,y,t;P)=\exp{px+Ey+\Omega t+\bfs\cdot\bft}
                \cdot \phi(\bfU x+\bfV y+ \bfW t+\bft,P)
\eqn\kpeigenvectors
$$
here the spectral curve $\Gamma$ is no more hyperelliptic,
and consequently the spectral parameter
$E$ becomes itself
a multi-valued function $E(P)$ of the point $P$ on the surface;
the functions $p(P),E(P),\Omega(P)$ are normalized by requiring
that they have {\it pure imaginary periods} along {\it all} cycles
$a_1,\ldots,a_g,b_1,\ldots,b_g$;
$\bfU,\bfV,\bfW$ are the {\it real} $2g$-dimensional vectors
of periods of the multi-valued functions $p(P),E(P),\Omega(P)$:
$$
\bfU={}^t(\oint_{b_1}dp,\ldots,\oint_{b_g}dp,
         -\oint_{a_1}dp,\ldots,-\oint_{a_g}dp),\qquad{\rm etc.};
\eqn\defbfu
$$
$t_1,\ldots,t_{2g}$ are auxiliary ``times'' needed for
further procedure of averaging; $s_1,\ldots,s_{2g}$
are the corresponding ``momenta'', not needed in what
follows;
$\phi$ is a {\it periodic function} with period 1 with respect to all of
the $2g$ variables $x_1,\ldots,x_g,y_1,\ldots,y_g$
($\abel$ is the Abel map):
$$
\phi(\pmatrix{\bfx\cr\bfy\cr};P)
        =c\, \exp{2\pi i \abel(P)\cdot\bfy}\cdot
        {\theta(\abel(P)+\bfx+B\bfy+\bfznot) \theta(\bfznot)\over
         \theta(\abel(P)+\bfznot) \theta(\bfx+B\bfy+\bfznot)}
\eqn\defphi
$$

We will also need the solutions to the adjoint system
$\psiad L = -\der{\psiad}{y}$,
$\psiad A = -\der{\psiad}{t}$
where differential operators {\it written on the left} should be intended
according to
$
\psiad (u \der{^j}{x^j}) \equiv (-\der{}{x})^j(\psiad u).
$
(formal integration by parts).
The left and right action differ only for a complete derivative:
$$
(\psiad L)\psi=\psiad L\psi+\der{}{x}(\psiad L^\one \psi)
                            +\der{{}^2}{x^2}(\psiad L^\two \psi)+\cdots
\eqn\revert
$$
as can readily be seen by repeated applications of the Leibnitz rule. Here
$L^{(r)}\equiv {(-1)^r\over r!}{d^r\over d(\d)^r} L$ (formal derivation with
respect to the symbol $\d$: for instance,
$A^\one=-3\d^2+{3\over 2} u$).

Solutions to the adjoint system can be written in the form
$
\psiad(x,y,t;P)=e^{-px-Ey-\Omega t-\bfs\cdot\bft}\cdot
                \phi^\dagger(-\bfU x-\bfV y- \bfW t-\bft,P)
$

The application of the Whitham method to equations of the KP type
is allowed by the possibility of averaging
identities by means of some ``ergodic theorem'': the average
of periodic functions
$\phi(t_1,\ldots,t_{2g};I_k)$ with period
1 with respect to all arguments is given by
$\<\phi\>\equiv\int\phi(\bft)d^{2g}t$, and for a generic vector $\bfU$
it coincides with the limit
$
\<\phi\>_x\equiv\lim_{x_0\rightarrow +\infty}{1\over 2x_0}
        \int_{-x_0}^{x_0}\phi(\bfU x)dx
$,
because the line $\bfU x$ winds densely on a $2g$-dimensional torus.
Note that the derivative of $\phi$ along any direction has zero
average value:
$$
\<\der{}{x}\phi(\bfU x)\>=\bfU\cdot\<\der{\phi}{\bft}\>=
        \sum_{j=1}^{2g}
        U_j\int_0^1\der{\phi}{t_j}dt_1\ldots dt_{2g}=0
\eqn\zeroaverage
$$
Note also that the functions
        $\der{\phi}{I_k}$
are again periodic,
because in our real normalization the periods have
the fixed value 1 (not depending on the $I_k$).

Solutions of the KP equation have the form
$$
u(x,y,t)=-2\dertwo\log\theta(\bfU x+\bfV y +\bfW t+{\bf z}_0|B)+C(B)
\eqn\kpnzonesolutions
$$
Thus, we look for a semiclassical approximation in the form
$$
u_0(x,y,t)  =-2\dertwo\log\theta({\bfS(X,Y,T)\over\eps}|I(X,Y,T))+C(X,Y,T),
$$
where
$\der{\bfS}{X}=\bfU$,
$\der{\bfS}{Y}=\bfV$,
$\der{\bfS}{T}=\bfW$,
$\psi_0=e^{{1\over\eps}\bfs\cdot\bfS} \phi({\bfS\over\eps})$.
The operators $L_0,A_0$ obtained substituting $u\mapsto u_0$
are taken as first terms of the asymptotic series
$$
A=A_0+\eps A_1+\cdots,\qquad L=L_0+\eps L_1+\cdots,
\eqn\laseries
$$

Let us introduce the notation
$$
\hat\der{}{\tau}\equiv
        \der{\bfI}{\tau}\cdot\der{}{\bfI}\equiv
        \sum_j \der{I_j}{\tau}\der{}{I_j};
\eqn\derivatives
$$
then the
substitution $u\mapsto u_0$ implies
$\der{}{x}\mapsto\der{}{x}+\eps\hat\der{}{X}$;
taking terms of order ${\cal O}(\eps)$ in \lax\ we get the linearized
equation
$$
\der{L_1}{t}-\der{A_1}{y}+[L_0,A_1]+[L_1,A_0]=K-F,
\eqn\linearized
$$
where $F$ is the term due to derivation with respect to slow variables:
this term must be adjusted in order to compensate for $K$.
We readily find for $F$ the form
$$
F=\hat\der{L}{T}-\hat\der{A}{Y}+(L^\one\hat\der{A}{X}-A^\one\hat\der{L}{X}).
\eqn\defef
$$
Now use
$\der{\psi}{y}=L\psi$,
$\der{\psiad}{y}=-\psiad L$,
$\der{\psi}{t}=A\psi$,
$\der{\psiad}{t}=-\psiad A$,
and \revert:
$$
\der{}{t}(\psiad L_1\psi)-\der{}{y}(\psiad A_1\psi)
=\psiad\Bigl(\der{L_1}{t}-\der{A_1}{y}
        +[L_0,A_1]+[L_1,A_0]\Bigr)\psi+\der{}{x}(\dots)
\eqn\totder
$$
Thus, the average of the left hand side of \linearized\ comes out to
be zero (being the average of a total derivative) and we obtain
the Whitham equations in the implicit form
$$
\<\psiad K\psi\>=\<\psiad F\psi\>
\eqn\whithamone
$$
Explicit computing of the Whitham term
$\<\psiad F\psi\>$
will give us the final form.
Take respectively

({\it i}) a curve $\bfI=\bfI(\tau)$ in the
space of parameters ($P=\const,\bft=\const$);

({\it ii}) a curve $P=P(\tau)$ moving the
point $P$ on the surface $\Gamma$ ($\bfI=\const, \bft=\const$);

({\it iii}) a curve $\bft=\bft(\tau)$ moving
only the ``times'' $t_i$ ($\bfI=\const,P=\const$).
\par\noindent
Correspondingly, we get $L(\tau),A(\tau)$, etc., and
$$
\eqalign{
\psi(\tau) &=\exp{p(\tau)x+E(\tau)y+\Omega(\tau)t+\bfs\cdot\bft}\cdot
                \phi(\bfU(\tau)x+\bfV(\tau)y+\bfW(\tau)t+\bft)  \cr
\psiad     &=\exp{-px-Ey-\Omega t-\bfs\cdot\bft}\cdot
                \phiad(-\bfU x-\bfV y-\bfW t-\bft)              \cr}
\eqn\variation
$$
Now compute
$\left. \der{}{\tau}\right|_{\tau=0}\psiad\psi(\tau)$ in all the three cases
and use a point to denote derivation with respect to $\tau$:
$$
\eqalign{
(i)\quad\derivative
        &=({\dot p} x+{\dot E} y+{\dot \Omega} t)\, \psiad\psi+
    ({\dot\bfU} x+{\dot\bfV} y+{\dot\bfW} t)
                \cdot\psiad\der{\psi}{\bft}+
        {\dot\bfI}\cdot\psiad\der{\psi}{\bfI},                  \cr
(ii)\quad\derivative &=(dp\; x+dE\; y+d\Omega\; t)\, \phiad\phi,\cr
(iii)\quad\derivative &=\bfs\cdot{\dot\bft}\, \phiad\phi+
                \phiad\der{\phi}{\bft}\cdot{\dot\bft};  \cr}
\eqn\tauderivative
$$
Using
        $\der{\psi}{t}=A\psi$,
        $\der{\psiad}=-\psiad A$
and \revert, see that
$$
\der{}{t}(\psiad\psi(\tau))
=\psiad(A(\tau)-A)\psi(\tau)
-\der{}{x}(\psiad A^\one\psi(\tau))+\der{^2}{x^2}(\ldots)
\eqn\identity
$$
Deriving the left-hand side with respect to $\tau$ and using $(i)$ we get
for instance
$$
\eqalign{
        {\left.\d\over\d\tau\right|_0}
        \der{}{t}(\psiad\psi(\tau))
        =&{\dot\Omega}\, \psiad\psi+{\dot\bfW}
                \cdot\psiad\der{\psi}{\bft}     \cr
        &+\{ ({\dot p} x+{\dot E} y +{\dot \Omega} t)\,
                \der{}{t}(\psiad\psi) \cr
        &+({\dot \bfU} x+{\dot\bfV} y+{\dot\bfW} t)\cdot
                \der{}{t}(\phiad\der{\phi}{\bft})
         +{\dot\bfI}
        \cdot\der{}{t}(\phiad\der{\phi}{\bfI})  \}      \cr
}
\eqn\example
$$
Now fix $x,y,t$ and average upon $d^{2g}t$: the terms in braces
are linear combinations of total derivatives with constant coefficients
and thus {\it vanish}, giving
$$
\<\left. \der{}{\tau}\right|_0\der{}{t}(\psiad\psi(\tau))\>=
{\dot \Omega}\, \<\psiad\psi\>+{\dot\bfW}\cdot\<\psiad\der{\psi}{\bft}\>;
\eqn\exampleone
$$
this passage contains the essence of the method of averaging. We
can now go on deriving both sides of \identity\ with
respect to $\tau$, applying $(i),(ii),(iii)$
and finding after averaging the following identities:
$$
\eqalign{
(i):\qquad & {\dot \Omega}\<\psiad\psi\>+
        {\dot\bfW}\cdot\<\psiad\der{\psi}{\bft}\>
        =\<\psiad\der{A}{\tau}\psi\>-{\dot p}\<\psiad A^\one\psi\>-
\bfU\cdot\<\phiad \hat A^\one\der{\phi}{\bft}\> \cr
(ii):\qquad &d\Omega \<\psiad\psi\>   =  -dp\<\psiad A^\one\psi\>  \hfill\cr
(iii):\qquad &0         =  \<\psiad\der{A}{t_j}\>  \hfill\cr}
\eqn\result
$$
Note that $\psiad\psi=\phiad\phi$, that we posed
$\hat A^\one=\exp{(-px-\cdots)}A^\one \exp{(px+\cdots)}$,
and that $(iii)$ implies
$$
\<\psiad\der{A}{\tau}\psi\>\equiv
  \<\psiad(\hat\der{A}{\tau} A
        +({\dot \bfU} x+{\dot\bfV} y+{\dot\bfW} t)
                \cdot\der{A}{\bft})\psi\>
        = \<\psiad\hat\der{A}{\tau} \psi\>
\eqn\hatderivative
$$
We get analogous identities for $E,\bfV, L$ if we derive instead of
\identity\ the identity
$$
\der{}{y}(\psiad\psi(\tau))
        =\psiad(L(\tau)-L)\psi(\tau)-
        \der{}{x}(\psiad L^\one\psi(\tau))+\der{^2}{x^2}(\ldots)
\eqn\identityone
$$
Now letting $\tau=Y,T$ (remember that $Y,T$ are independent of $y,t$), we
can rewrite \result\ and the analogous identities for $L$ as
$$
\eqalign{
-\<\psiad\hat\der{A}{y}\psi\> &=-\der{\Omega}{Y}\<\psiad\psi\>
        -\der{\bfW}{Y}\cdot\<\phiad\der{\phi}{\bft}\>
        -\der{p}{Y}\<\psiad \hat A^\one\psi\>
        -\der{\bfU}{Y}\cdot\<\phiad\der{\phi}{\bft}\>   \cr
\<\psiad\hat\der{L}{t}\psi\> &=\der{E}{T}\<\psiad\psi\>
        +\der{\bfV}{T}\cdot\<\phiad\der{\phi}{\bft}\>
        +\der{p}{T}\<\psiad \hat L^\one\psi\>
        +\der{\bfU}{T}\cdot\<\phiad\der{\phi}{\bft}\>   \cr}
\eqn\eqnone
$$
The last term we need comes from the identity
$$
\der{}{t}(\psiad L^\one \psi(\tau))-\der{}{y}(\psiad A^\one\psi(\tau))=
  \psiad[L^\one(A(\tau)-A)-A^\one(L(\tau)-L)]\psi(\tau)+
\der{}{x}(\ldots)
\eqn\identitytwo
$$
which, after putting $\tau=X$ and averaging, gives
$$
\<\psiad(L^\one\der{A}{X}-A^\one\der{L}{X})\psi\>=
\der{\Omega}{X}\<\psiad L^\one\psi\>- \der{E}{X}\<\psiad A^\one\psi\>
+\der{\bfW}{X} \cdot\<\psiad \hat L^\one\der{\phi}{\bft}\>
-\der{\bfW}{X} \cdot\<\phiad \hat A^\one \der{\phi}{\bft}\>
\eqn\eqntwo
$$
Summing up \eqnone\ and \eqntwo, and using the compatibility conditions
$$
\der{\bfU}{Y}=\der{\bfW}{X},\qquad
\der{\bfU}{T}=\der{\bfW}{X},\qquad
\der{\bfV}{T}=\der{\bfW}{Y}
\eqn\compatibility
$$
we get
$$
\<\psiad F\psi\>
  =(\der{\Omega}{Y}-\der{E}{T})\<\psiad\psi\>
+(\der{\Omega}{X}-\der{p}{T})\<\psiad L^\one\psi\>
+(\der{p}{Y}-\der{E}{X})\<\psiad A^\one\psi\>
\eqn\basta
$$
using $(ii)$ from \result\ we can rewrite the Whitham equations in
the final form
$$
(\der{\Omega}{Y}-\der{E}{T})dp+
(\der{p}{T}-\der{\Omega}{X})dE+
(\der{E}{X}-\der{p}{Y})d\Omega=
{\<\psiad K\psi\> \over \<\psiad\psi\>}dp.
\eqn\lastwhitham
$$

\section{Whitham equations for the stationary and evolutive KdV}

The KdV is a particular case of the KP equation.
The solutions of the KP equation correspond to generic (non-hyperelliptic,
that is not of the form $w^2=p(E)$) Riemann surfaces $\Gamma$: in this
case the function $E(P), P\in\Gamma$
itself is no more one-valued and its differential
$dE$ has non-zero periods $V_j={1\over 2\pi i}\oint_{b_j} dE$.
The solutions have the form \kpnzonesolutions.
When $\Gamma$ is hyperelliptic the differential
$dE$ is exact, $\bfV=0$ and the dependence on
$y$ disappears, giving solutions to the KdV equation.
For KdV the Whitham equations have the form
$$
(\der{p}{T}-\der{\Omega}{X})={\<\psiad K\psi\>\over\<\psiad\psi\>}{dp\over dE}
\eqn\kdvwhitham
$$
Solutions to the {\it stationary} KdV equation, which is equivalent to the
linear system
$L\psi =E\psi$,
$A\psi =w(E)\, \psi$,
come out when $d\Omega$ too is exact, and this is true for
$\Omega(E)=2 w(E)dE=\sqrt{4 E^3-g_2 E-g_3}\, dE$; the
corresponding Whitham equation gives
$$
\der{w}{X}\, dE= {\<\psiad K\psi\>\over\<\psiad\psi\>} dp
\eqn\stationarywhitham
$$
where $dp$ is normalized with ${\rm Im}\oint dp=0$.

\section{Integrability of the Whitham equations for K=0}

For KdV the Whitham equations have the form \kdvwhitham.
In this case, the parameters $I_j$ of the preceding section are simply the
branch points ${\bf E}={}^t(E_1,\ldots,E_{2g+1})$ of the spectral curve.
For $K=0$ this comes out as
$$
\der{p}{T}=\der{\Omega}{X}
\eqn\zerok
$$
Krichever has shown that
\zerok\ has solutions
$E_1(X,T),\ldots,E_{2g+1}(X,T)$ given implicitly by the
conditions
$$
{d\Lambda\over dp}(\bfE(X,T))|_{E=E_j(X,T)}+X+
        T\, {d\Omega\over dp}(\bfE(X,T))|_{E=E_j(X,T)}=0,\qquad
        j=1,\ldots,2g+1,
\eqn\whithamsolutionforkdv
$$
where $d\Lambda$ is an arbitrary differential with possibly discontinuities
and singularities not depending on $X,T$. Analogous solutions exist for
the KP case (see Ref. \averagingmethod).

In order to see it, consider
that if the function
        $S(X,T)=\int_\infty^P d\Lambda(X,T)+Xdp(X,T)+Td\Omega(X,T)$
is such that
        $\der{S}{X}=p,\der{S}{T}=\Omega$,
then\zerok\ is automatically satisfied.
Now, \whithamsolutionforkdv\ is equivalent to
        $dS|_{E=E_j(X,T)}=0$
for
        $j=1,\ldots,2g+1$;
the form $\der{}{X} dS=dp+(\der{}{X} d\Lambda+X\der{}{X} dp+
T\der{}{X}d\Omega)$
has the same normalization as $dp$, the same
singularity at the infinity and is holomorphic
everywhere else, except, generally speaking, in the points
$E_j$, because {\it e.g.}
$
\der{}{X}{dE\over \wtorus}=
-{1\over 2}{(\ldots)dE\over[(E-E_1)(E-E_2)(E-E_3)]^{3/2}}+\cdots
$
cease to be holomorphic in these points. However, thanks to
$dS|_{E=E_j}=0$
we have $dS=\sqrt{E-E_j}dE+\calO(E-E_j)^{3/2}$ in a
neighborhood of
$E_j$, and $\der{}{X} dS=-\der{E_j}{X}{dE\over 2\sqrt{E-E_j}}+\cdots=
        -\der{E_j}{X}{2 u du\over 2 u+\cdots}$
comes out to be non-singular in $E_j$ too: thus
$\der{}{X} dS$ coincides with $dp$, and similarly
$\der{}{T}dS=d\Omega$.

\section{Integrability of the stationary Whitham equation for $K=1$}

The equation
        $\der{w}{X}dE=dp$
 can be integrated in the following way
(see Ref. \novkrich): take
        $k=\sqrt{E}$, $dE=2k\, dk$,
        $w=\sqrt{E^{2g+1}+c_1 E_{2g-1}+\cdots+c_{2g}}=
           k^{2g+1}+{2g+1\over 2} T_{2g+1} k^{2g-1}
          +{2g-1\over 2} T_{2g-1} k^{2g-3}+\cdots
          +{3\over 2} T_3 k+{x\over 2} {1\over k}
          +\calO({1\over k^2})$
and require
        ${\rm Im} \oint_{a_1}w\, dE=h_i=\const$,
        ${\rm Im} \oint_{b_1}w\, dE=h'_i=\const$,
        $i=1,\ldots,g$.
The first condition gives $c_1,\ldots,c_g$ as algebraic functions
of
        $T_{2g+1},\ldots,T_3$;
the second condition fixes also
        $c_{g+1},\ldots,c_{2g}$
as transcendental functions of
        $T_{2g+1},\ldots,T_3,X$.
Moreover, the first condition gives
        $\der{}{X}w\, dE\simeq d\sqrt{E}\simeq dp$,
while the second assures that both sides have the same
normalization: this means that
        $\der{}{X}w\, dE=dp$,
as desired.
(Introducing the functions
        $\Omega_j\simeq k^j+\calO (1)$,
        $E\rightarrow\infty$,
it is easy to see that
        $\der{}{T_j}w\, dE = d\Omega_j$,
thus giving also
        $\der{\Omega_i}{T_j}=\der{\Omega_j}{T_i}$).

\refout
\end